\documentclass[10pt,journal]{IEEEtran}

\usepackage{amsmath,amssymb,amsfonts,bm}
\usepackage{algorithm}
\usepackage{algorithmic}
\usepackage{array}
\usepackage{booktabs}
\usepackage{cite}
\usepackage{graphicx}
\usepackage{subfig}

\usepackage{xcolor}

\usepackage{tikz}
\usepackage{url}
\usetikzlibrary{arrows.meta,positioning,fit,calc}

\graphicspath{{../}{figs/}{./}}

\newcommand{\Uu}{\mathcal{U}_{\mathrm{u}}}
\newcommand{\Ue}{\mathcal{U}_{\mathrm{e}}}
\newcommand{\Ucal}{\mathcal{U}}
\newcommand{\Kcal}{\mathcal{K}}
\newcommand{\Expect}{\mathbb{E}}
\newcommand{\Reals}{\mathbb{R}}

\newcommand{\indicator}{\mathbf{1}}

\DeclareMathOperator*{\argmax}{arg\,max}
\DeclareMathOperator*{\argmin}{arg\,min}

\begin{document}

\title{Agentic Quantum Deep Reinforcement Learning for RAN Slicing}

\author{
    Tingnan Bao,~\IEEEmembership{Senior Member,~IEEE},
    Medhat Elsayed,
    Pedro Enrique Iturria-Rivera,
    Yigit Ozcan,
    Majid Bavand,
    Melike Erol-Kantarci,~\IEEEmembership{Fellow,~IEEE}
    \thanks{T. Bao and M. Erol-Kantarci are with the School of Electrical Engineering and Computer Science, University of Ottawa, Ottawa, ON, Canada (e-mail: tbao@uottawa.ca, melike.erolkantarci@uottawa.ca).}
    \thanks{M. Elsayed, M. Bavand, P. E. Iturria-Rivera, and Y. Ozcan are with Ericsson, Canada (e-mail: \{medhat.elsayed, majid.bavand, pedro.iturria.rivera, yigit.ozcan\}@ericsson.com).}
}

\maketitle

\begin{abstract}
Radio access network (RAN) slicing enables ultra-reliable low-latency communications (URLLC) and enhanced mobile broadband (eMBB) services to share radio resources, but their requirements create a challenging reliability--throughput tradeoff. URLLC requires low-latency and reliable packet delivery, whereas eMBB targets high sustained throughput. This paper considers downlink URLLC/eMBB RAN slicing and formulates it as a queue-aware long-term eMBB throughput maximization problem subject to URLLC delay-violation, physical resource block (PRB) exclusivity, and slice-budget constraints. To solve this problem, we propose agentic quantum deep reinforcement learning (Agentic-QDRL), a two-time-scale framework that combines agentic slice-level resource control with quantum-enhanced PRB scheduling. At the slow time scale, a perceive--memory--act--reflect (PMAR) controller adapts the resource shares of URLLC and eMBB slices. At the fast time scale, a compact variational quantum circuit (VQC)-based QDRL scheduler performs PRB allocation under the current slice configuration. A feasibility projection and a safety fallback mechanism is further introduced to satisfy scheduling constraints and reduce URLLC deadline violations. Simulation results under different eMBB traffic loads show that Agentic-QDRL improves eMBB throughput, reduces eMBB queue buildup, and maintains URLLC delay reliability compared with classical DRL and heuristic baselines.
\end{abstract}

\begin{IEEEkeywords}
Agentic AI, quantum deep reinforcement learning, RAN slicing, URLLC/eMBB coexistence, variational quantum circuit.
\end{IEEEkeywords}

\section{Introduction}

Fifth-generation (5G) and sixth-generation (6G) mobile networks are expected to support diverse services with very different performance requirements~\cite{sun2025advancing}. Network slicing addresses this heterogeneity by partitioning a shared physical infrastructure into multiple logical networks designed for different service classes~\cite{li2025incremental}. In the radio access network (RAN), slicing is implemented by allocating finite radio resources, especially physical resource blocks (PRBs), among different slices~\cite{zhao2025adaslicing}. Given the resource share of each slice, the RAN scheduler assigns PRBs to users within each slice at each scheduling interval. Since all slices share the same PRB pool, allocating more resources to one slice reduces the resources available to the others. Therefore, RAN slicing performance depends on how the scheduler balances different slice demands under time-varying traffic and channel conditions. This balance becomes difficult when services with conflicting requirements compete for the same radio resources.

Ultra-reliable low-latency communications (URLLC) and enhanced mobile broadband (eMBB) are two major service classes in RAN slicing with conflicting service requirements. URLLC services require low-latency and highly reliable packet delivery, whereas eMBB services require high and sustained throughput. This makes URLLC/eMBB coexistence a challenging resource-allocation problem. To address this problem, the authors in~\cite{Alsenwi2019} proposed a risk-sensitive URLLC resource-allocation scheme for URLLC/eMBB coexistence, using conditional value-at-risk (CVaR) to protect vulnerable eMBB users and a chance constraint to ensure URLLC reliability. The authors in~\cite{Anand2020} developed joint eMBB/URLLC schedulers where eMBB allocation considers the expected rate loss caused by minislot-level URLLC superposition or puncturing, while URLLC demands are served immediately. In~\cite{han2020qos}, the authors studied differential quality-of-service (QoS)-oriented URLLC scheduling under a first-in-first-out (FIFO) queuing model and used a cross-layer Markov-chain model to analyze average transmission power and delay. More recently, the authors in~\cite{Taskou2024E2E} studied end-to-end resource slicing for URLLC/eMBB coexistence in 5G-Advanced/6G networks. The authors in~\cite{Liu2024RSMA} applied rate-splitting multiple access (RSMA) to uplink network slicing and showed that RSMA can improve the achievable rate region under heterogeneous slice requirements. These studies show that URLLC protection and eMBB throughput maximization involve a fundamental tradeoff over shared radio resources. This tradeoff is difficult to manage because wireless channel conditions and traffic demand vary over time, which motivates learning-based resource control.

Deep reinforcement learning (DRL) has become a promising approach for dynamic radio resource allocation because it can learn long-term control policies from interactions with the network environment~\cite{azimi2021energy,filali2024open}. The authors in~\cite{benmadani2025deep} proposed a DRL-based dynamic downlink scheduling algorithm for real-time LTE traffic and extended it to 5G RAN slicing with a focus on eMBB slices. The authors in~\cite{qiao2025resource} proposed a multi-timescale Open RAN (O-RAN) slicing framework in which hierarchical DRL jointly controls inter-slice and intra-slice radio and computing resources under delay and reliability requirements. The authors in~\cite{Filali2022SDN} proposed a two-time-scale software-defined networking (SDN)-based RAN slicing mechanism that uses an exponential-weight algorithm for exploration and exploitation at large-time-scale resource allocation and multi-agent deep Q-learning (DQL) for short-time-scale scheduling. The authors in~\cite{Zhang2024CRSDRL} proposed constrained risk-sensitive DRL for eMBB/URLLC joint scheduling, using CVaR to reduce extreme scheduling risks while mitigating the impact of URLLC puncturing on eMBB users. A broader survey of reinforcement learning for radio resource management in RAN slicing was provided in~\cite{Zangooei2023RLSurvey}. Despite this progress, DRL approaches for URLLC/eMBB slicing still face large action spaces and real-time scheduling requirements. These challenges motivate compact policy models for learning-based resource control.

To this end, quantum reinforcement learning (QRL) has recently emerged as a complementary way to build compact policy models~\cite{tran2025quantum,puspitasari2025quantum}. Since fault-tolerant quantum computers are not yet widely available, most practical quantum learning studies focus on noisy intermediate-scale quantum (NISQ) devices~\cite{preskill2018quantum}. In this setting, a variational quantum circuit (VQC) is commonly used as a trainable quantum model because it combines shallow parameterized quantum gates with classical optimization. The authors in~\cite{Chen2020VQCDRL} proposed a VQC-based DQL framework that uses quantum circuits to approximate the Q-value function and reduces the number of trainable parameters through quantum information encoding. The authors in~\cite{Kolle2024QuantumA2C} proposed a VQC-based advantage actor-critic framework and showed that hybrid quantum-classical actor-critic designs can improve performance on benchmark control tasks. The authors in~\cite{Ansere2024QDRL} proposed a quantum-empowered DRL (Qe-DRL) approach for mobile-edge-computing (MEC)-based Internet of Things (IoT) offloading and showed improved learning speed and energy-efficiency performance over classical benchmarks. The authors in~\cite{Zhang2025VQRLVEC} proposed a tensor-network-preprocessed quantum DRL (QDRL) algorithm for high-dimensional vehicular edge computing resource allocation and reported faster convergence with fewer quantum resources than benchmark methods. A recent survey in~\cite{puspitasari2025QDRLSurvey} summarized QDRL methods for wireless resource allocation. These studies show the potential of QRL for compact and efficient decision making. However, applying VQC-based policies to URLLC/eMBB RAN slicing remains underexplored.

In parallel, agentic AI has emerged as a design principle for autonomous 6G networks, where agents perceive network states, use stored experience, take actions, and reflect on outcomes~\cite{chergui2026tutorial,xiao2025toward}. The authors in~\cite{xiao2025toward} proposed AgentNet, a framework that uses generative foundation models as agents to support interaction, collaborative learning, and knowledge transfer among AI agents in 6G networks. The authors in~\cite{Zheng2026AgenticDRL} reviewed advanced DRL as a core enabler of agentic AI for wireless communications, highlighting its role in autonomous decision-making, reasoning, and long-term adaptation. The authors in~\cite{Ferrag2026Agents6G} proposed an agentic AI-native 6G architecture where bounded and policy-governed agents operate within a semantic control plane above deterministic network infrastructure. The authors in~\cite{Habib2025AgenticRANSlicing} proposed an agentic RAN-slicing framework in which a large language model (LLM) translates operator intents into actionable goals and hierarchical controllers coordinate inter-slice, intra-slice, and self-healing agents. These works show that agentic control can support autonomous network management through a perceive--memory--act--reflect (PMAR) process. However, existing studies still treat learning-based scheduling, quantum policy design, and agentic control largely in isolation. In particular, agentic slicing frameworks mainly focus on high-level orchestration, while the integration of QDRL policies with agentic budget control and constraint-aware scheduling for URLLC/eMBB RAN slicing remains underexplored.

To address this gap, this paper proposes an Agentic-QDRL framework for URLLC/eMBB RAN slicing. The framework combines agentic slice-level resource control with quantum-enhanced PRB scheduling. At the slow time scale, the agentic controller adjusts the resource shares of the URLLC and eMBB slices. At the fast time scale, the QDRL scheduler makes PRB allocation decisions under the current slice-resource configuration. Feasibility projection and safety fallback mechanisms are further used to satisfy scheduling constraints and protect URLLC reliability. The main contributions are summarized as follows:

\begin{itemize}
\item We formulate downlink URLLC/eMBB RAN slicing as a queue-aware long-term eMBB throughput maximization problem subject to URLLC delay-violation, PRB exclusivity, and slice-budget constraints.

\item We propose a two-time-scale Agentic-QDRL architecture that couples a PMAR budget controller with a compact VQC-based QDRL scheduler.

\item We introduce a feasibility projection and a safety fallback mechanism to satisfy PRB exclusivity and slice-budget constraints while reducing URLLC deadline violations.

\item We evaluate the proposed framework under multiple eMBB traffic loads and ablation settings, demonstrating improved eMBB throughput while maintaining URLLC delay reliability compared with classical DRL and heuristic baselines.
\end{itemize}

\begin{figure}[!t]
\centering
\includegraphics[width=\columnwidth]{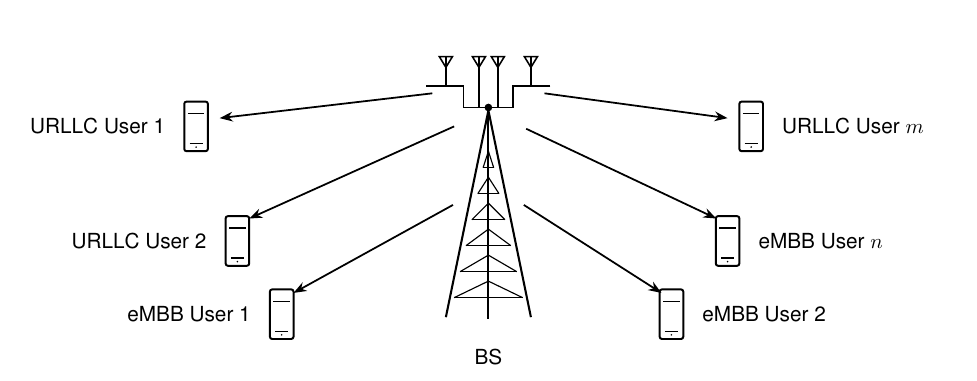}
\caption{System model of the single-cell downlink URLLC/eMBB RAN slicing.}
\label{fig:system_model}
\end{figure}
\section{System Model and Problem Formulation}
\label{sec:system_model}

We consider a single-cell downlink OFDMA RAN slicing scenario, where a base station (BS) serves a set of UEs $\Ucal$ over a shared pool of $K$ PRBs, as shown in Fig.~\ref{fig:system_model}. The UE set is partitioned into a URLLC user set $\Uu$ and an eMBB user set $\Ue$, with $\Ucal=\Uu\cup\Ue$ and $\Uu\cap\Ue=\emptyset$. Time is divided into scheduling intervals indexed by $t\in\{0,1,\ldots\}$, each with duration $\Delta t$. The PRBs are indexed by $\Kcal=\{1,\ldots,K\}$. At each scheduling interval, the BS schedules PRBs among the UEs by deciding the PRB-to-UE assignment. We define $x_{u,k}(t)\in\{0,1\}$ as the corresponding scheduling variable, where $x_{u,k}(t)=1$ if PRB $k$ is assigned to UE $u$ in scheduling interval $t$, and $x_{u,k}(t)=0$ otherwise. Due to OFDMA orthogonality, at most one UE can occupy each PRB; thus, the assignment variables satisfy
\begin{equation}
\sum_{u\in\Ucal}x_{u,k}(t)\leq 1,\qquad \forall k\in\Kcal,\;\forall t ,
\label{eq:exclusive}
\end{equation}

Building upon this PRB-level exclusivity, RAN slicing introduces a two-level resource allocation. At the slice level, the BS determines per-scheduling-interval PRB budgets for the URLLC and eMBB slices, and these budgets are kept fixed over a slicing interval consisting of $T_o$ consecutive scheduling intervals. Within each scheduling interval, the scheduler assigns PRBs from the corresponding slice budget to individual UEs. Let $w$ index the slicing interval, with $\mathcal{T}_w=\{wT_o,\ldots,(w+1)T_o-1\}$ denoting the set of scheduling intervals covered by the same slice budgets. During slicing interval $w$, the PRB budgets allocated to the URLLC and eMBB slices, denoted by $B_{\mathrm{u}}^{(w)}$ and $B_{\mathrm{e}}^{(w)}$, satisfy
\begin{equation}
B_{\mathrm{u}}^{(w)}+B_{\mathrm{e}}^{(w)}\leq K  ,\qquad B_{\mathrm{u}}^{(w)},\;B_{\mathrm{e}}^{(w)}\in\{0,1,\ldots,K\}.
\label{eq:budget_sum}
\end{equation}
For each $t\in\mathcal{T}_w$, the executed PRB assignment must satisfy
\begin{align}
\sum_{u\in\Uu}\sum_{k\in\Kcal}x_{u,k}(t)&\leq B_{\mathrm{u}}^{(w)},
\label{eq:budget_u}\\
\sum_{u\in\Ue}\sum_{k\in\Kcal}x_{u,k}(t)&\leq B_{\mathrm{e}}^{(w)}.
\label{eq:budget_e}
\end{align}

\subsection{Channel and Service Model}
The wireless channel between the BS and each UE comprises large-scale fading from path loss and shadowing, and small-scale fading from multipath propagation. The large-scale gain between the BS and user $u$ follows a log-distance path-loss model:
\begin{equation}
\beta_u(t)=C_0\!\left(\frac{d_u(t)}{d_0}\right)^{-\alpha}10^{-\xi_u(t)/10},
\label{eq:pathloss}
\end{equation}
where $C_0$ is the reference gain, $d_u(t)$ denotes the distance between BS and UE $u$, $d_0$ is the reference distance, $\alpha$ is the path-loss exponent, and $\xi_u(t)\sim\mathcal N(0,\sigma_{\mathrm{sh}}^2)$ is the shadowing term in dB. The small-scale coefficient follows a Rician fading model:
\begin{equation}
\tilde h_{u,k}(t)=
\sqrt{\frac{\kappa}{\kappa+1}}\,h_{u,k}^{\mathrm{LoS}}(t)+
\sqrt{\frac{1}{\kappa+1}}\,h_{u,k}^{\mathrm{NLoS}}(t),
\label{eq:rician}
\end{equation}
where $h_{u,k}^{\mathrm{LoS}}(t)$ is a deterministic unit-power LoS component, $h_{u,k}^{\mathrm{NLoS}}(t)\sim\mathcal{CN}(0,1)$ is the random NLoS component, $\kappa$ is the Rician factor, and $\Expect[|\tilde h_{u,k}(t)|^2]=1$.
Combining the large- and small-scale effects, the downlink channel between the BS and UE $u$ on PRB $k$ in scheduling interval $t$ is
\begin{equation}
h_{u,k}(t)=\sqrt{\beta_u(t)}\,\tilde h_{u,k}(t).
\label{eq:channel}
\end{equation}

When PRB $k$ is scheduled to UE $u$, the received signal is
\begin{equation}
y_{u,k}(t)=\sqrt{p_k}\,h_{u,k}(t)s_{u,k}(t)+n_{u,k}(t),
\label{eq:rx_signal}
\end{equation}
where $p_k=P_{\max}/K$ denotes the fixed transmit power on each PRB under equal power allocation, and $P_{\max}$ is the total BS transmit-power budget,  $s_{u,k}(t)$ is the transmitted symbol with $\Expect[|s_{u,k}(t)|^2]=1$, and $n_{u,k}(t)\sim\mathcal{CN}(0,\sigma^2)$ is modeled as additive white Gaussian noise (AWGN) with zero mean and variance $\sigma^2$, which represents the thermal noise power. Since the considered single-cell OFDMA model assigns each PRB to at most one UE and neglects inter-cell interference, the received signal-to-noise ratio (SNR) is
\begin{equation}
\gamma_{u,k}(t)=
\frac{p_k\beta_u(t)|\tilde h_{u,k}(t)|^2}{\sigma^2},
\label{eq:snr}
\end{equation}
As such, the achievable PRB rate is
\begin{equation}
r_{u,k}(t)=W_{\mathrm{PRB}}\log_2\!\big(1+\gamma_{u,k}(t)\big),
\label{eq:rate_prb}
\end{equation}
where $W_{\mathrm{PRB}}$ is the PRB bandwidth.
The aggregate scheduled rate and the corresponding potential service bits of user $u$ are
\begin{equation}
\widehat R_u(t)=\sum_{k\in\Kcal}x_{u,k}(t)r_{u,k}(t),\qquad
\widehat S_u(t)=\widehat R_u(t)\Delta t .
\label{eq:potential_service}
\end{equation}

\subsection{Traffic and Queuing Model}
We model downlink traffic using a UE-associated queue for each user, which stores data waiting to be scheduled for transmission to the corresponding UE. At the beginning of scheduling interval $t$, let $Q_u(t)$ denote the queue length of UE $u$ in bits. Since actual delivery cannot exceed the available backlog, the served bits and delivered throughput of UE $u$ are
\begin{equation}
S_u(t)=\min\{Q_u(t),\widehat S_u(t)\},\qquad
R_u(t)=\frac{S_u(t)}{\Delta t}.
\label{eq:actual_service}
\end{equation}
Here, $S_u(t)$ is measured in bits and $R_u(t)$ is measured in bits/s. Let $A_u(t)$ denote the traffic volume admitted for UE $u$ during scheduling interval $t$, also measured in bits. We adopt a post-service arrival convention, under which $S_u(t)$ is first served from the backlog $Q_u(t)$ and $A_u(t)$ is then admitted to the queue. Consequently, newly admitted data starts with zero waiting time in the next state, and the queue evolves as
\begin{equation}
Q_u(t+1)=\big[Q_u(t)-S_u(t)\big]^+ + A_u(t),
\label{eq:queue}
\end{equation}
where $[\cdot]^+=\max(\cdot,0)$.

Since URLLC and eMBB have different traffic characteristics, their arrivals are modeled separately. For URLLC traffic, we model bursty arrivals using a Bernoulli process with sporadic packet generation~\cite{Alsenwi2019}:
\begin{equation}
A_u(t)=
\begin{cases}
L_u, & \text{w.p. }p_u^{\mathrm{arr}},\\
0, & \text{w.p. }1-p_u^{\mathrm{arr}},
\end{cases}
\qquad u\in\Uu ,
\label{eq:urllc_arr}
\end{equation}
where $L_u$ is the URLLC packet size and $p_u^{\mathrm{arr}}$ is the arrival probability.  For eMBB traffic, we use a Poisson bit-arrival model to approximate persistent downlink traffic \cite{Anand2020}:
\begin{equation}
A_u(t)\sim\mathrm{Poisson}(\lambda_e),\qquad u\in\Ue ,
\label{eq:embb_arr}
\end{equation}
where $\lambda_e$ is the mean eMBB traffic volume in bits per UE per scheduling
interval.

\subsection{URLLC Delay Reliability and eMBB Throughput Metrics}
The queuing model above is used to define the slice-level service-level agreement (SLA) metrics. For URLLC, reliability is deadline-based, meaning that a packet is considered successfully served only if it is delivered before the latency deadline. Hence, for each URLLC user $u\in\Uu$, we track a head-of-line (HoL) delay state $D_u(t)$, defined as the waiting time of the oldest URLLC data present at the beginning of scheduling interval $t$. Under first-in-first-out (FIFO) service and the post-service arrival convention in \eqref{eq:queue}, residual backlog after service keeps the HoL state active and increases its delay by $\Delta t$. If no residual backlog remains after service, the HoL state is reset to zero, including the case where new data is admitted at the end of the interval. The HoL delay evolves as
\begin{equation}
D_u(t+1)=
\begin{cases}
D_u(t)+\Delta t, & \big[Q_u(t)-S_u(t)\big]^+>0,\\
0, & \text{otherwise}.
\end{cases}
\label{eq:hol}
\end{equation}
Given the URLLC latency deadline $\tau_u$, the instantaneous delay-violation indicator $v_u(t)$ and slice-level URLLC delay-violation rate $V(t)$ are
\begin{equation}
v_u(t)=\indicator\{D_u(t)>\tau_u\},\qquad
V(t)=\frac{1}{|\Uu|}\sum_{u\in\Uu}v_u(t),
\label{eq:violation}
\end{equation}
respectively. For eMBB, the primary service metric is sustained broadband throughput. Accordingly, the instantaneous eMBB throughput utility, measured in bits/s, is
\begin{equation}
U(t)=\sum_{u\in\Ue}R_u(t).
\label{eq:utility}
\end{equation}
To characterize long-term performance under stochastic channel and traffic dynamics, we define the long-term expected eMBB throughput $\bar U$ and the long-term expected URLLC delay-violation rate $\bar V$ as
\begin{equation}
\begin{aligned}
\bar U=\liminf_{T\to\infty}\frac{1}{T}\sum_{t=0}^{T-1}\Expect[U(t)], \qquad
\bar V=\limsup_{T\to\infty}\frac{1}{T}\sum_{t=0}^{T-1}\Expect[V(t)],
\end{aligned}
\label{eq:long_term}
\end{equation}
respectively. 

\subsection{Problem Formulation}
Our objective is to determine the slice-level PRB budgets $\{B_{\mathrm{u}}^{(w)},B_{\mathrm{e}}^{(w)}\}$ and the scheduling-interval PRB-to-UE assignment variables $\{x_{u,k}(t)\}$ that maximize the long-term eMBB throughput while satisfying the URLLC delay-reliability requirement. The optimization problem is formulated as
\begin{equation}
\begin{aligned}
\mathrm{(P1)}\quad
\max_{\substack{\{x_{u,k}(t)\}\\ \{B_{\mathrm{u}}^{(w)},B_{\mathrm{e}}^{(w)}\}}}\;\;& \bar U\\
\mathrm{s.t.}\quad
&\mathrm{C1}: \bar V\leq\varepsilon, \\
&\mathrm{C2}: \eqref{eq:exclusive}, \\
&\mathrm{C3}:  \eqref{eq:budget_sum}\text{--}\eqref{eq:budget_e}, \\
&\mathrm{C4}:  x_{u,k}(t)\in\{0,1\},\qquad \forall u,k,t,
\end{aligned}
\label{eq:problem}
\end{equation}
where $\varepsilon$ denotes the maximum URLLC delay-violation rate. Constraint $\mathrm{C1}$ enforces the URLLC delay-reliability requirement, $\mathrm{C2}$ captures OFDMA orthogonality, ensuring that each PRB is assigned to at most one UE in each scheduling interval, $\mathrm{C3}$ imposes slice-level PRB budget feasibility, including the total PRB budget and the per-slice URLLC/eMBB resource limits within each slicing interval, and $\mathrm{C4}$ enforces binary PRB-to-UE scheduling decisions.

Problem~(P1) is NP-hard since it contains the 0-1 knapsack problem as a deterministic special case. Fix the slice budgets and assume that the URLLC slice already satisfies C1. For the remaining eMBB slice in one scheduling interval, let each UE $u$ have one predefined transmission option with value $a_u$ and PRB cost $c_u$. Selecting options under budget $B_{\mathrm{e}}$ gives $\max_{\{y_u\}}\sum_{u\in\Ue}a_u y_u$ subject to $\sum_{u\in\Ue}c_u y_u\leq B_{\mathrm{e}}$ and $y_u\in\{0,1\}$, which is exactly 0-1 knapsack. Since this restricted instance is NP-hard, the original problem with PRB-level assignments, adaptive budgets, queue evolution, and long-term delay-reliability constraints is also NP-hard. As such, we adopt the Agentic-QDRL framework to learn scalable slicing and scheduling policies while enforcing hard PRB assignment and slice-budget constraints.

\section{Proposed Agentic-QDRL Framework}

\begin{figure*}[!t]
\centering
\includegraphics[width=\textwidth]{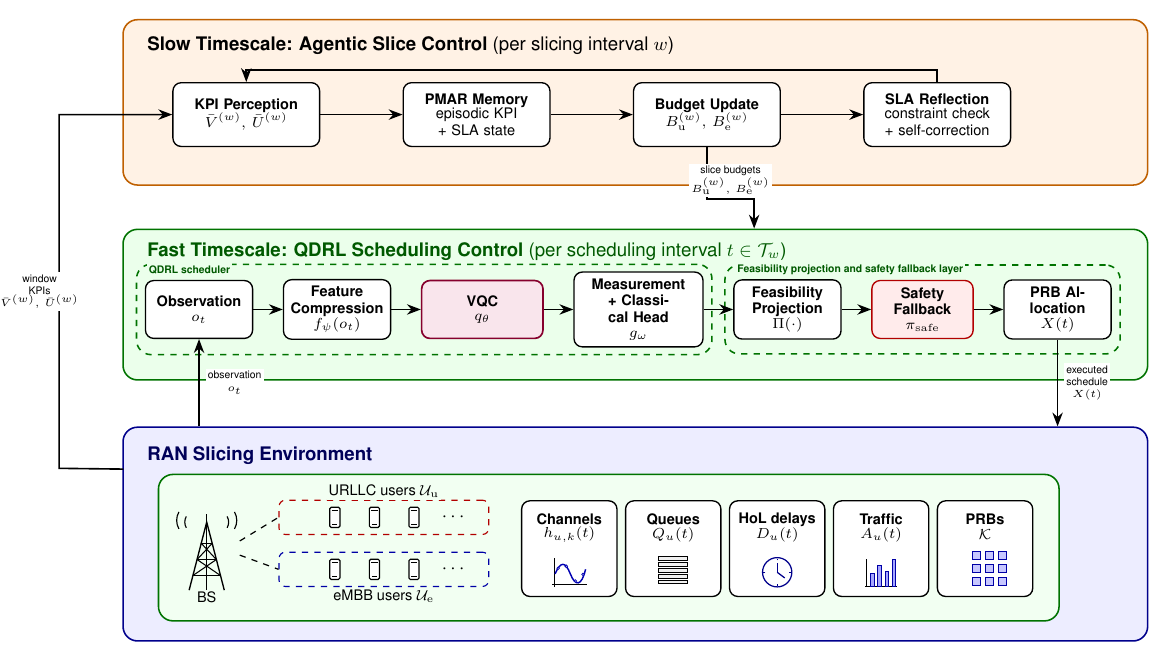}
\caption{Agentic-QDRL architecture for URLLC/eMBB RAN slicing.}
\label{fig:framework}
\end{figure*}

In this section, we present the proposed Agentic-QDRL framework for delay-reliability-aware URLLC/eMBB RAN slicing. As illustrated in Fig.~\ref{fig:framework}, the design follows a two-time-scale control architecture. The slow time scale governs slice-level resource adaptation according to aggregate SLA and network KPIs, while the fast time scale performs PRB scheduling under the current slice budgets. The two levels interact through a closed feedback loop: the slow controller provides slice budgets to the scheduler, the scheduler executes feasible PRB allocations, and the resulting service and queue dynamics are fed back for subsequent budget adaptation. The following subsections describe the slow agentic budget controller, the QDRL scheduler, the feasibility and safety mechanisms, and the learning procedure.

\subsection{Agentic PMAR Budget Control}
The slow controller follows the PMAR loop over slicing intervals. Its role is not to schedule individual PRBs, but to regulate the resource envelope within which the fast scheduler operates. At the end of each slicing interval, the controller perceives slice-level KPIs, stores the resulting state in structured memory, acts by updating the URLLC/eMBB PRB budgets, and reflects on SLA risk by adjusting the safety state used by the fast scheduler.

\subsubsection{Perceive}
During slicing interval $w$, the same slice budgets are applied to all scheduling intervals in $\mathcal{T}_w$. After the interval finishes, the controller aggregates the fast scheduling outcomes into interval-level delay-reliability and throughput KPIs:
\begin{equation}
\begin{aligned}
\bar U^{(w)}=\frac{1}{T_o}\sum_{t\in\mathcal{T}_w}U(t), \qquad
\bar V^{(w)}=\frac{1}{T_o}\sum_{t\in\mathcal{T}_w}V(t).
\end{aligned}
\label{eq:perceive_kpis}
\end{equation}

\subsubsection{Memory}
The PMAR controller maintains two types of structured memory. The episodic memory records what happened in recent slicing intervals, while the constraint memory stores the rules that govern budget adaptation. Let $\mathcal{H}_w$ denote the set of retained slicing intervals, with $|\mathcal{H}_w|\leq H$. For each retained interval $i\in\mathcal{H}_w$, the episodic memory stores a record
\begin{equation}
\begin{aligned}
\mathcal{M}_{\mathrm{epi}}^{(i)}
=\{\bar U^{(i)},\bar V^{(i)},
B_{\mathrm{u}}^{(i)},B_{\mathrm{e}}^{(i)}\}.
\end{aligned}
\label{eq:episodic_memory}
\end{equation}
Thus, the episodic memory tells the controller how previous budget decisions affected throughput and realized delay-reliability. In contrast, the constraint memory stores the persistent control rules
\begin{equation}
\mathcal{M}_{\mathrm{con}}^{(w)}
=\{\varepsilon,\tau_u,\tau_{\mathrm{safe}}^{(w)},
B_{\mathrm{u}}^{\min},B_{\mathrm{u}}^{\max}\},
\label{eq:constraint_memory}
\end{equation}
where $\tau_{\mathrm{safe}}^{(w)}$ is the current safety threshold maintained by the slow controller.

\subsubsection{Act}
The action of the slow controller is the update of the slice-level PRB budgets. For this update, the controller extracts reliability, fallback, and backlog feedback signals from the current slicing interval and the episodic memory:
\begin{equation}
\begin{aligned}
\bar V_{\mathrm{mem}}^{(w)}
&=\frac{1}{|\mathcal{H}_w|}\sum_{i\in\mathcal{H}_w}
\bar V^{(i)}, \qquad
\bar F^{(w)} =\frac{1}{T_o}\sum_{t\in\mathcal{T}_w}F(t),\\
\bar Q_{\mathrm{e}}^{(w)}
&=\frac{1}{T_o|\Ue|}\sum_{t\in\mathcal{T}_w}\sum_{u\in\Ue}Q_u(t).
\end{aligned}
\label{eq:act_auxiliary}
\end{equation}
where $F(t)$ is equal to one if the safety fallback is triggered in scheduling interval $t$, and zero otherwise. To prevent the slow controller from overestimating the safety of the learned scheduler, the act stage also forms a diagnostic nominal-violation signal. Let $X_{\mathrm{nom}}(t)$ denote the projected allocation proposed by the learned scheduler before safety fallback. We define
\begin{equation}
\bar V_{\mathrm{nom}}^{(w)}
=\frac{1}{T_o}\sum_{t\in\mathcal{T}_w}V_{\mathrm{nom}}(t),
\label{eq:nominal_violation}
\end{equation}
where $V_{\mathrm{nom}}(t)$ is the violation rate that would have occurred under $X_{\mathrm{nom}}(t)$ before fallback correction, while $V(t)$ remains the realized violation rate after possible fallback correction. If fallback is not triggered, $V_{\mathrm{nom}}(t)=V(t)$. The controller then forms a composite budget error
\begin{equation}
e^{(w)}=\mathrm{clip}\!\left(e_{\mathrm{SLA}}^{(w)}+e_{\mathrm{fb}}^{(w)}
+e_{\mathrm{gap}}^{(w)}+e_{\mathrm{embb}}^{(w)},-e_{\mathrm{clip}},e_{\mathrm{clip}}\right),
\label{eq:budget_error}
\end{equation}
with
\begin{equation}
\begin{aligned}
e_{\mathrm{SLA}}^{(w)}=\bar V_{\mathrm{mem}}^{(w)}-\varepsilon, \qquad
e_{\mathrm{fb}}^{(w)}=\lambda_{\mathrm{fb}}\big[\bar F^{(w)}-\beta_{\mathrm{fb}}\big]^+,\\
e_{\mathrm{gap}}^{(w)}=\lambda_{\mathrm{gap}}\big[\bar V_{\mathrm{nom}}^{(w)}-\bar V^{(w)}\big]^+, 
e_{\mathrm{embb}}^{(w)}=-\lambda_q\left[\frac{\bar Q_{\mathrm{e}}^{(w)}}{Q_{\mathrm{th}}}-1\right]^+ .
\end{aligned}
\end{equation}
where $\lambda_{\mathrm{fb}}$, $\lambda_{\mathrm{gap}}$, and $\lambda_q$ are nonnegative control gains for fallback usage, nominal-realized violation gap, and eMBB backlog pressure, respectively, $\beta_{\mathrm{fb}}$ is the tolerated fallback-trigger rate, $Q_{\mathrm{th}}$ is the eMBB backlog threshold, and $\mathrm{clip}(\cdot,-e_{\mathrm{clip}},e_{\mathrm{clip}})$ bounds the composite budget error within $[-e_{\mathrm{clip}},e_{\mathrm{clip}}]$. The first three terms increase the URLLC budget when the delay-reliability target is violated, fallback is used too often, or safety correction hides nominal learned-scheduler violations. The last term acts as an eMBB backlog-pressure regularizer motivated by Lyapunov drift-based stochastic network control~\cite{neely2010stochastic}. It decreases the URLLC budget when the eMBB backlog persistently exceeds $Q_{\mathrm{th}}$.

To avoid reducing the URLLC slice below a load- and safety-aware minimum, the controller uses an adaptive lower projection bound
\begin{equation}
B_{\mathrm{u}}^{\mathrm{floor},(w)}
=
\Big[
B_{\mathrm{u},0}
+\Delta_{\mathrm{safe}}^{(w)}
\Big]_{B_{\mathrm{u}}^{\min}}^{B_{\mathrm{u}}^{\max}},
\label{eq:budget_floor}
\end{equation}
where $B_{\mathrm{u},0}\in[B_{\mathrm{u}}^{\min},B_{\mathrm{u}}^{\max}]$ denotes the load-based lower budget determined by the offered URLLC traffic, and $\Delta_{\mathrm{safe}}^{(w)}\geq 0$ is a safety-aware increment determined by recent fallback usage and the nominal-realized delay-violation gap. The URLLC budget is then updated by
\begin{equation}
B_{\mathrm{u}}^{(w+1)}=
\Big[B_{\mathrm{u}}^{(w)}+\eta_B e^{(w)}\Big]_{B_{\mathrm{u}}^{\mathrm{floor},(w)}}^{B_{\mathrm{u}}^{\max}},
\label{eq:budget_urllc_update}
\end{equation}
where $[x]_a^b=\min\{\max\{x,a\},b\}$ denotes projection onto $[a,b]$. The eMBB budget takes the residual resources
\begin{equation}
B_{\mathrm{e}}^{(w+1)}=K-B_{\mathrm{u}}^{(w+1)}.
 \label{eq:budget_embb_update}
\end{equation}

\subsubsection{Reflect}
The reflect stage checks whether the current operating point violates the URLLC delay-reliability target and updates the safety state used by the fast scheduler. In particular, the controller computes a safety margin from the memory-averaged violation and then tightens the safety threshold as
\begin{equation}
\begin{aligned}
\Delta\tau^{(w)}
&=\Delta\tau_0+\rho\big[\bar V_{\mathrm{mem}}^{(w)}-\varepsilon\big]^+,\\
\tau_{\mathrm{safe}}^{(w+1)}
&=\Big[\tau_u-\Delta\tau^{(w)}\Big]_{\tau_{\min}}^{\tau_u}.
\end{aligned}
\label{eq:reflect}
\end{equation}
where $\Delta\tau_0$ is the base safety margin, $\rho$ controls the reflection strength, and $\tau_{\min}$ is the minimum admissible safety threshold. Thus, when the nominal scheduler repeatedly approaches an unsafe URLLC operating point, the fallback mechanism is activated earlier in subsequent scheduling intervals. This reflection step closes the slow-time-scale PMAR loop and provides a safety-aware state for the next slicing interval.

\subsection{QDRL Scheduler}
The fast scheduler operates at each scheduling interval and maps the current RAN observation and slice budgets to UE priority scores. Instead of learning binary PRB assignments directly, the VQC actor learns a continuous score-space policy. The resulting scores are first projected into a feasible candidate PRB allocation that satisfies the resource constraints, and this candidate allocation can be overridden by the fallback mechanism when the URLLC operating point becomes unsafe.

\subsubsection{Observation}
At each scheduling interval, the scheduler observes compact cross-layer features
\begin{equation}
o_t=\big(\{\bar\gamma_u(t)\},\{Q_u(t)\},\{\bar r_u(t)\},\{D_u(t)\},B_{\mathrm{u}}^{(w)},B_{\mathrm{e}}^{(w)}\big),
\label{eq:scheduler_observation}
\end{equation}
where $\bar\gamma_u(t)=\frac{1}{K}\sum_{k\in\Kcal}\gamma_{u,k}(t)$ and $\bar r_u(t)=\frac{1}{K}\sum_{k\in\Kcal}r_{u,k}(t)$ denote the average SNR and average achievable PRB rate of UE $u$, respectively. 

\subsubsection{Feature Compression}
Since the observation dimension grows with the number of UEs, a classical encoder first compresses $o_t$ into an $n$-dimensional vector matched to the number of qubits:
\begin{equation}
\mathbf f_t=f_{\psi}(o_t)\in\Reals^n,\qquad
\mathbf a_t=\pi\tanh(\mathbf f_t),
\label{eq:feature_compression}
\end{equation}
where $\psi$ denotes the encoder parameters and $\mathbf a_t$ is the bounded angle vector used for quantum encoding.

\subsubsection{Variational Quantum Actor}
\begin{figure}[!t]
\centering
\includegraphics[width=\columnwidth]{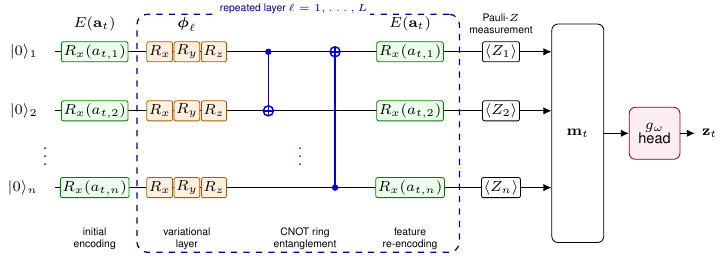}
\caption{VQC architecture used by the QDRL scheduler.}
\label{fig:vqc}
\end{figure}

As shown in Fig.~\ref{fig:vqc}, the VQC acts as a compact quantum analog of a neural network that maps the compressed feature vector $\mathbf a_t$ to quantum measurement features through a shallow data-reuploading circuit composed of parameterized single-qubit rotations and entangling gates, making it suitable for NISQ devices.

\textit{Quantum state initialization and initial data encoding.}
The $n$-qubit register is initialized as $|0\rangle^{\otimes n}$, where $|\cdot\rangle^{\otimes n}$ denotes the tensor product of $n$ identical single-qubit states. The bounded feature vector $\mathbf a_t$ is first encoded by angle rotations,
\begin{equation}
E(\mathbf a_t)=\bigotimes_{j=1}^{n}R_x(a_{t,j}),
\label{eq:angle_encoding}
\end{equation}
where $R_x(\cdot)$ is a rotation around the Pauli-$X$ axis. This encoding maps each compressed classical feature to the rotation angle of one qubit.

\textit{Variational layer.}
At layer $\ell$, each qubit is transformed by trainable single-qubit rotations,
\begin{equation}
V_\ell(\bm\phi_\ell)=
\bigotimes_{j=1}^{n}R_z(\phi_{\ell,j,3})R_y(\phi_{\ell,j,2})R_x(\phi_{\ell,j,1}),
\label{eq:vqc_layer}
\end{equation}
where $\bm\phi_\ell=\{\phi_{\ell,j,1},\phi_{\ell,j,2},\phi_{\ell,j,3}\}_{j=1}^n$ denotes the trainable parameters in layer $\ell$.

\textit{Entangling layer.}
After the trainable rotations, a ring of CNOT gates entangles neighboring qubits:
\begin{equation}
C_{\mathrm{ring}}
=\mathrm{CNOT}_{n,1}\mathrm{CNOT}_{n-1,n}\cdots
\mathrm{CNOT}_{2,3}\mathrm{CNOT}_{1,2}.
\label{eq:cnot_ring}
\end{equation}

\textit{Feature re-uploading.}
The same angle encoding $E(\mathbf a_t)$ is re-applied after the entangling layer, allowing the classical features to interact repeatedly with the trainable quantum parameters. The complete VQC unitary is
\begin{equation}
U(\mathbf a_t,\bm\theta)=
\prod_{\ell=1}^{L}
\left[E(\mathbf a_t)C_{\mathrm{ring}}V_\ell(\bm\phi_\ell)\right],
\label{eq:vqc_unitary}
\end{equation}
where $\bm\theta=\{\bm\phi_\ell\}_{\ell=1}^L$ collects the trainable VQC parameters. In each repeated block, the operations are applied to the current state in the order $V_\ell(\bm\phi_\ell)$, $C_{\mathrm{ring}}$, and $E(\mathbf a_t)$. The final circuit state is
\begin{equation}
|\psi(\mathbf a_t,\bm\theta)\rangle
=U(\mathbf a_t,\bm\theta)E(\mathbf a_t)|0\rangle^{\otimes n}.
\label{eq:vqc_state}
\end{equation}

\subsubsection{Measurement and Classical Head}
The VQC output is read out through Pauli-$Z$ expectation values:
\begin{equation}
m_j(t)=
\langle\psi(\mathbf a_t,\bm\theta)|\hat Z_j|\psi(\mathbf a_t,\bm\theta)\rangle,
\qquad j=1,\ldots,n .
\label{eq:vqc_measurement}
\end{equation}
The measurement vector $\mathbf m_t=[m_1(t),\ldots,m_n(t)]^\mathsf{T}$ is then mapped by a classical output head to UE priority scores:
\begin{equation}
\mathbf z_t=g_\omega(\mathbf m_t)=W_{\mathrm{out}}\mathbf m_t+\mathbf b_{\mathrm{out}}
\in\Reals^{|\Ucal|},
\label{eq:score_head}
\end{equation}
where $\omega=\{W_{\mathrm{out}},\mathbf b_{\mathrm{out}}\}$ denotes the classical head parameters. The score $z_u(t)$ represents the scheduler's preference for UE $u$ under the current observation and slice budgets.

\subsubsection{Score-Space Exploration}
During training, Gaussian exploration is added in the score space:
\begin{equation}
\tilde{\mathbf z}_t=\mathbf z_t+\bm\xi_t,\qquad
\bm\xi_t\sim\mathcal N(\mathbf 0,\sigma_\xi^2\mathbf I).
\label{eq:score_exploration}
\end{equation}
The exploratory vector $\tilde{\mathbf z}_t$ is used as input to the feasibility projection.

\subsection{Feasibility Projection and Safety Fallback}
The VQC actor produces continuous UE scores, whereas the RAN scheduler must execute binary PRB assignments that satisfy OFDMA exclusivity, slice-budget limits, and URLLC delay-safety requirements. Therefore, Agentic-QDRL places two deterministic post-processing modules after the learned score generator. The feasibility projection converts the score vector into a constraint-satisfying nominal schedule, while the safety fallback replaces this nominal schedule only when URLLC HoL delays approach the safety threshold.

\subsubsection{Feasibility Projection}
Given the exploratory score vector $\tilde{\mathbf z}_t$ and slice budgets $(B_{\mathrm{u}}^{(w)},B_{\mathrm{e}}^{(w)})$, the projection $\Pi(\cdot)$ maps continuous UE preferences to a binary nominal allocation $X_{\mathrm{nom}}(t)$. It first partitions the available PRBs into disjoint URLLC and eMBB budget sets, denoted by $\Kcal_{\mathrm{u}}^{(w)}$ and $\Kcal_{\mathrm{e}}^{(w)}$, with
\begin{equation}
|\Kcal_{\mathrm{u}}^{(w)}|\leq B_{\mathrm{u}}^{(w)},\quad
|\Kcal_{\mathrm{e}}^{(w)}|\leq B_{\mathrm{e}}^{(w)},\quad
\Kcal_{\mathrm{u}}^{(w)}\cap\Kcal_{\mathrm{e}}^{(w)}=\emptyset .
\label{eq:projection_partition}
\end{equation}
Within each slice $s\in\{\mathrm{u},\mathrm{e}\}$, each PRB is then assigned to the UE with the largest score-rate product:
\begin{equation}
u_k^\star=\argmax_{u\in\Ucal_s}\tilde z_u(t) r_{u,k}(t), \quad k\in\Kcal_s^{(w)}.
\label{eq:projection_rule}
\end{equation}
For the URLLC slice, the maximization is taken over backlogged URLLC users
whenever such users exist. If all URLLC queues are empty, the unused URLLC PRBs
are released to the eMBB slice to avoid resource idling. For eMBB, the projection
follows the score-rate rule under the persistent-traffic model, while the actual
served bits remain queue-limited by \eqref{eq:actual_service}. The corresponding
binary entry is set as $x_{u_k^\star,k}(t)=1$, while all other UE entries on PRB
$k$ remain zero.

The resulting nominal schedule is denoted by
\begin{equation}
X_{\mathrm{nom}}(t)=
\Pi\!\left(\tilde{\mathbf z}_t;
B_{\mathrm{u}}^{(w)},B_{\mathrm{e}}^{(w)}\right).
\label{eq:x_nominal}
\end{equation}
Because $\Kcal_{\mathrm{u}}^{(w)}$ and $\Kcal_{\mathrm{e}}^{(w)}$ are disjoint and \eqref{eq:projection_rule} selects at most one UE for each PRB, $X_{\mathrm{nom}}(t)$ satisfies OFDMA exclusivity and slice-budget constraints by construction.

\subsubsection{Safety Fallback}
Although the projection guarantees resource feasibility, it does not by itself guarantee that urgent URLLC packets are prioritized when their HoL delays become close to the latency deadline. We therefore use a deterministic fallback policy as a safety layer around the learned scheduler. The fallback is triggered when
\begin{equation}
\max_{u\in\Uu}D_u(t)>\tau_{\mathrm{safe}}^{(w)} .
\label{eq:fallback_trigger}
\end{equation}
If \eqref{eq:fallback_trigger} is not satisfied, the executed schedule is $X(t)=X_{\mathrm{nom}}(t)$. Otherwise, the fallback policy $\pi_{\mathrm{safe}}$ replaces the nominal schedule. 

For the URLLC slice, the safety fallback $\pi_{\mathrm{safe}}$ follows an earliest-deadline-first (EDF) rule:
\begin{equation}
u_k^\star=\argmin_{u\in\Uu,\;Q_u(t)>0}\big(\tau_u-D_u(t)\big),
\qquad k\in\Kcal_{\mathrm{u}}^{(w)} .
\label{eq:edf_fallback}
\end{equation}
For the eMBB slice, the safety fallback $\pi_{\mathrm{safe}}$ uses proportional-fair (PF) scheduling:
\begin{equation}
u_k^\star=\argmax_{u\in\Ue}\frac{r_{u,k}(t)}{\bar R_u^{\mathrm{ema}}(t)},
\qquad k\in\Kcal_{\mathrm{e}}^{(w)} ,
\label{eq:pf_fallback}
\end{equation}
where $\bar R_u^{\mathrm{ema}}(t)$ is the moving average served throughput of eMBB UE $u$. 

In this way, the learned policy remains responsible for normal scheduling decisions, while the fallback provides explicit protection when the URLLC operating point becomes unsafe.

\subsection{Learning Algorithm}
The overall learning procedure couples the slow time-scale agentic PMAR-based slice controller with the fast time-scale QDRL scheduler. During a slicing interval, the current budgets and safety threshold are fixed, and the QDRL scheduler repeatedly produces score-space actions, projects them to feasible PRB allocations, and collects replay transitions. After the interval ends, the PMAR controller aggregates the resulting KPIs, updates its memory, and computes the slice budgets and safety threshold for the next interval.

The QDRL actor is trained with a classical critic $V_\phi(o_t)$. The replay buffer stores score-space transitions formed by the observation $o_t$, the exploratory score action $\tilde{\mathbf z}_t$, the reward used for actor--critic learning, and the next observation used in the learning update. Let $U_{\mathrm{tr}}(t)$, $V_{\mathrm{tr}}(t)$, and $o_{t+1}^{\mathrm{tr}}$ denote the eMBB throughput, URLLC delay-violation rate, and next observation associated with this replay transition. If fallback is inactive, these quantities are obtained from the executed schedule $X(t)$. If fallback is triggered, they are evaluated counterfactually under the nominal schedule $X_{\mathrm{nom}}(t)$, while the reward still includes a fallback penalty so that unsafe score actions are discouraged. The learning reward is
\begin{equation}
r_t^{\mathrm{RL}}=U_{\mathrm{tr}}(t)-\lambda_v[V_{\mathrm{tr}}(t)-\varepsilon]^+
-\lambda_{\mathrm{fb}}^{\mathrm{RL}}\indicator\{\pi_{\mathrm{safe}}\ \mathrm{triggered}\}.
\label{eq:reward}
\end{equation}
The critic is updated by minimizing the squared TD error
\begin{equation}
\delta_t=r_t^{\mathrm{RL}}+\chi V_{\phi^-}(o_{t+1}^{\mathrm{tr}})-V_\phi(o_t),
\label{eq:td_error}
\end{equation}
where $\chi \in (0,1)$ is the discount factor and $V_{\phi^-}$ is a slowly updated target critic used for stabilizing bootstrapping. The critic loss is
\begin{equation}
\mathcal L_{\mathrm{c}}(\phi)=
\Expect_{\mathcal D}\!\left[\delta_t^2\right],
\label{eq:critic_loss}
\end{equation}
and the critic parameters are updated by
\begin{equation}
\phi \leftarrow \phi-\eta_{\mathrm{c}}\nabla_{\phi}\mathcal L_{\mathrm{c}}(\phi),
\qquad
\phi^- \leftarrow \tau_{\mathrm{tar}}\phi+(1-\tau_{\mathrm{tar}})\phi^- ,
\label{eq:critic_update}
\end{equation}
where $\eta_{\mathrm{c}}$ is the critic learning rate and $\tau_{\mathrm{tar}} \in (0,1)$ is the target-network smoothing factor.
The actor is updated through the score-space Gaussian policy gradient:
\begin{equation}
\nabla_\Theta J(\Theta)\approx
\Expect\!\left[\nabla_\Theta\log\pi_\Theta(\tilde{\mathbf z}_t|o_t)\delta_t\right],
\end{equation}
and the actor parameters are updated by gradient ascent as
\begin{equation}
\Theta \leftarrow \Theta+\eta_{\mathrm a}\nabla_\Theta J(\Theta),
\label{eq:actor_update}
\end{equation}
where $\eta_{\mathrm a}$ is the actor learning rate. Here $\Theta=\{\psi,\bm\theta,\omega\}$ collects all trainable actor parameters: $\psi$ for the feature encoder, $\bm\theta=\{\bm\phi_\ell\}_{\ell=1}^{L}$ for the VQC variational rotations, and $\omega$ for the classical output head. The score-space policy is modeled as
\begin{equation}
\pi_\Theta(\tilde{\mathbf z}_t|o_t)
=\mathcal N(\tilde{\mathbf z}_t;\mathbf z_t,\sigma_\xi^2\mathbf I),
\label{eq:score_policy}
\end{equation}
where $\mathbf z_t$ is the mean score vector produced by the hybrid VQC actor in \eqref{eq:score_head}.
Gradients associated with the classical encoder and output head are obtained by standard automatic differentiation. For the VQC parameters $\bm\theta$, the required circuit derivatives can be evaluated by the parameter-shift rule~\cite{mitarai2018,schuld2019eval}, which obtains each partial derivative from two evaluations of the same circuit with the corresponding gate parameter shifted positively and negatively. This keeps the hybrid policy-gradient update compatible with NISQ hardware. Therefore, the complete two-time-scale Agentic-QDRL learning procedure is summarized in Alg.~\ref{alg:agentic_qdrl_learning}.

\begin{algorithm}[!t]
\caption{Agentic-QDRL Framework for RAN Slicing}
\label{alg:agentic_qdrl_learning}
\begin{algorithmic}[1]
\STATE Initialize actor $\Theta$, critic $\phi$, target critic $\phi^-$, budgets $(B_{\mathrm{u}}^{(0)},B_{\mathrm{e}}^{(0)})$, safety threshold $\tau_{\mathrm{safe}}^{(0)}$, PMAR memories $\mathcal M_{\mathrm{epi}}$ and $\mathcal M_{\mathrm{con}}$, and replay buffer $\mathcal D$.
\FOR{slicing interval $w=0,1,\ldots$}
  \STATE \textit{/* Slow: Agentic budget update */}
  \IF{$w>0$}
    \STATE \textbf{Perceive:} aggregate interval KPIs $\bar U^{(w-1)}$ and $\bar V^{(w-1)}$ from $\mathcal{T}_{w-1}$ by \eqref{eq:perceive_kpis}.
    \STATE \textbf{Memory:} update episodic memory by \eqref{eq:episodic_memory} and retain constraint memory by \eqref{eq:constraint_memory}.
    \STATE \textbf{Act:} compute auxiliary control signals by \eqref{eq:act_auxiliary} and the nominal violation signal by \eqref{eq:nominal_violation}; form the budget error by \eqref{eq:budget_error}, compute the URLLC floor budget by \eqref{eq:budget_floor}, and update $(B_{\mathrm{u}}^{(w)},B_{\mathrm{e}}^{(w)})$ by \eqref{eq:budget_urllc_update}--\eqref{eq:budget_embb_update}.
    \STATE \textbf{Reflect:} update the safety threshold $\tau_{\mathrm{safe}}^{(w)}$ by \eqref{eq:reflect}.
  \ENDIF
  \STATE Keep the current PMAR state $(B_{\mathrm{u}}^{(w)},B_{\mathrm{e}}^{(w)},\tau_{\mathrm{safe}}^{(w)})$ fixed for all scheduling intervals $t\in\mathcal{T}_w$.
  \STATE \textit{/* Fast: QDRL scheduling */}
  \FOR{scheduling interval $t\in\mathcal{T}_w$}
    \STATE Observe $o_t$ by \eqref{eq:scheduler_observation}, compute VQC scores $\mathbf z_t$ by \eqref{eq:feature_compression}--\eqref{eq:score_head}, and sample exploratory scores $\tilde{\mathbf z}_t$ by \eqref{eq:score_policy}.
    \STATE \textit{/* Feasibility projection and safety fallback */}
    \STATE Compute nominal allocation $X_{\mathrm{nom}}(t)$ by the projection in \eqref{eq:x_nominal}.
    \IF{$\max_{u\in\Uu}D_u(t)>\tau_{\mathrm{safe}}^{(w)}$}
      \STATE Form the replay transition from the nominal allocation $X_{\mathrm{nom}}(t)$.
      \STATE Execute the safety fallback $X(t)=\pi_{\mathrm{safe}}(t)$.
    \ELSE
      \STATE Execute $X(t)=X_{\mathrm{nom}}(t)$.
    \ENDIF
    \STATE Store $(o_t,\tilde{\mathbf z}_t,r_t^{\mathrm{RL}},o_{t+1}^{\mathrm{tr}})$ in $\mathcal D$ and update the critic and actor by \eqref{eq:critic_update} and \eqref{eq:actor_update}.
  \ENDFOR
  \STATE Store the outcomes collected over $\mathcal{T}_w$ for the next PMAR update.
\ENDFOR
\end{algorithmic}
\end{algorithm}

\section{Simulation Results}
In this section, we evaluate the performance of the proposed Agentic-QDRL framework through simulations. The system and traffic parameters are summarized in Tab.~\ref{tab:system_params}. We consider a single-cell downlink OFDMA RAN slicing scenario with $K=52$ PRBs, serving $|\Uu|=32$ URLLC users and $|\Ue|=15$ eMBB users. UEs are uniformly distributed within a single cell of radius $R_{\mathrm{cell}}=289$ m. URLLC traffic follows the Bernoulli packet-arrival model with a common arrival probability $p_u^{\mathrm{arr}}=p^{\mathrm{arr}}=0.78$, while the eMBB load is $\lambda_e\in\{500,2000,4000,8000\}$ bits per scheduling interval.  For the proposed Agentic-QDRL framework, the learning hyperparameters are provided in Tab. \ref{tab:learning_params}.  The proposed quantum actor uses a 6-qubit, 2-layer VQC.

\begin{table}[!t]
\centering
\caption{System and Traffic Parameters}
\label{tab:system_params}
\footnotesize
\begin{tabular}{@{}p{0.44\columnwidth}p{0.20\columnwidth}p{0.26\columnwidth}@{}}
\toprule
Parameter & Symbol & Value \\
\midrule
Number of PRBs & $K$ & 52 \\
PRB bandwidth & $W_{\mathrm{PRB}}$ & 180 kHz \\
Scheduling interval & $\Delta t$ & 1 ms \\
Total BS transmit power  & $P_{\max}$ & 46 dBm \\
Noise power & $\sigma^2$ & $-121.45$ dBm \\
Path-loss exponent & $\alpha$ & 3.5 \\
Shadowing std. & $\sigma_{\mathrm{sh}}$ & 6 dB \\
Cell radius & $R_{\mathrm{cell}}$ & 289 m \\
URLLC/eMBB users & $|\Uu|/|\Ue|$ & $32/15$ \\
URLLC deadline & $\tau_u$ & 1 ms \\
SLA target & $\varepsilon$ & 0.02 \\
URLLC packet size & $L_u$ & 1024 bits \\
URLLC arrival prob. & $p^{\mathrm{arr}}$ & 0.78 \\
eMBB load & $\lambda_e$ & $0.5$--$8$ kbits/interval \\
\bottomrule
\end{tabular}
\end{table}

\begin{table}[!t]
\centering
\caption{Agentic-QDRL Learning Parameters}
\label{tab:learning_params}
\footnotesize
\begin{tabular}{@{}p{0.42\columnwidth}p{0.24\columnwidth}p{0.24\columnwidth}@{}}
\toprule
Parameter & Symbol & Value \\
\midrule
\multicolumn{3}{@{}l}{\textit{Agentic PMAR controller}} \\
\midrule
Slicing interval & $T_o$ & 50 intervals \\
Memory horizon & $H$ & 20 windows \\
Budget bounds & $(B_{\mathrm{u}}^{\min},B_{\mathrm{u}}^{\max})$ & $(5,40)$ \\
Budget step size & $\eta_B$ & 50 \\
Fallback gain; threshold & $\lambda_{\mathrm{fb}};\beta_{\mathrm{fb}}$ & $0.02;0.20$ \\
Nominal-gap gain & $\lambda_{\mathrm{gap}}$ & 2.0 \\
eMBB backpressure & $\lambda_q;Q_{\mathrm{th}}$ & $0.005;2\times10^5$ bits \\
Error clipping bound & $e_{\mathrm{clip}}$ & 0.05 \\
Initial safety threshold & $\tau_{\mathrm{safe}}^{(0)}$ & 0.5 ms \\
Safety margin; rate & $\Delta\tau_0;\rho$ & $0.2$ ms; $0.5$ \\
\midrule
\multicolumn{3}{@{}l}{\textit{QDRL scheduler}} \\
\midrule
Qubits; VQC depth & $n;L$ & $6;2$ \\
Discount factor & $\chi$ & 0.99 \\
Actor learning rate & $\eta_{\mathrm a}$ & $5\times10^{-4}$ \\
Critic learning rate & $\eta_{\mathrm c}$ & $10^{-3}$ \\
Replay buffer size & $|\mathcal D|$ & $5\times10^4$ \\
Batch size & $B_{\mathrm{batch}}$ & 128 \\
Soft target update & $\tau_{\mathrm{tar}}$ & 0.005 \\
Exploration noise & $\sigma_\xi$ & $0.1$ \\
Violation/fallback penalties & $\lambda_v;\lambda_{\mathrm{fb}}^{\mathrm{RL}}$ & $5\times10^7;10^7$ \\
\bottomrule
\end{tabular}
\end{table}

To demonstrate the performance of the proposed Agentic-QDRL framework, we compare against the following baselines and ablation variants:
\begin{itemize}
\item \textbf{Agentic-DRL}: A classical DRL baseline that replaces the VQC actor with a two-hidden-layer MLP, while keeping the PMAR budget controller, feasibility projection, and safety fallback unchanged.
\item \textbf{Heuristic+Agentic}: A deterministic scheduling baseline that uses EDF for URLLC users and PF scheduling for eMBB users under the same agentic budget adaptation.
\item \textbf{QDRL (w/o Agentic)}: A QDRL baseline that performs per-scheduling-interval PRB scheduling with the VQC-based QDRL scheduler under fixed URLLC/eMBB slice budgets, while using the same feasibility projection and safety fallback.
\item \textbf{Agentic-QDRL (w/o Fallback)}: An Agentic-QDRL variant that removes the safety fallback, while retaining the VQC actor, PMAR budget controller, and feasibility projection.
\end{itemize}

\begin{figure}[!t]
\centering
\includegraphics[width=\columnwidth]{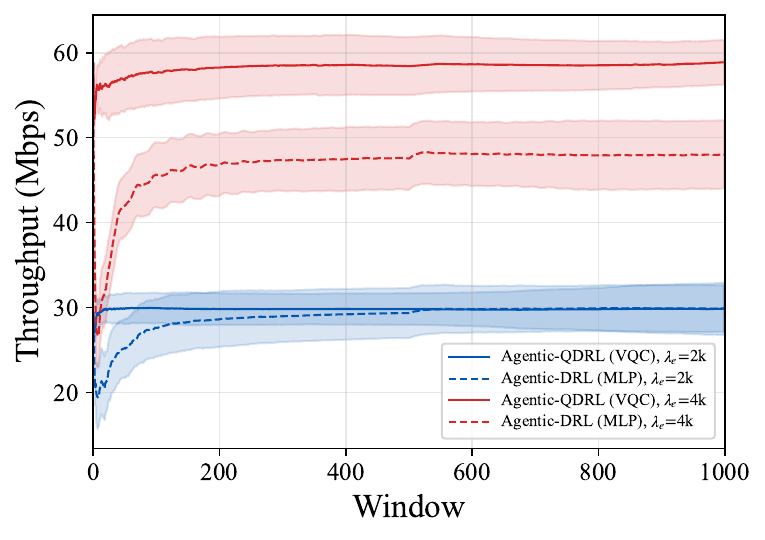}
\caption{Convergence of eMBB throughput for Agentic-QDRL (VQC) and Agentic-DRL (MLP) under different eMBB traffic loads $\lambda_e$.} 
\label{fig:conv_throughput}
\end{figure}

Fig.~\ref{fig:conv_throughput} plots the convergence of eMBB throughput over slicing windows for Agentic-QDRL (VQC) and Agentic-DRL (MLP) under two different eMBB traffic loads, $\lambda_e=2000$ and $\lambda_e=4000$ bits per scheduling interval. We observe that both learning-based methods reach a steady operating point after the initial training phase, while Agentic-QDRL settles within the first few tens of windows and Agentic-DRL needs several hundred windows to climb to a comparable level. This behavior is expected because the actor--critic scheduler improves its score-based PRB allocation policy over slicing windows, while the PMAR controller progressively stabilizes the slice-level resource budgets. We also see that a higher eMBB traffic load leads to a higher achieved eMBB throughput, because more persistent eMBB queues provide sufficient backlogged traffic to better utilize the available PRBs. Moreover, Agentic-QDRL achieves a stronger steady-state throughput than Agentic-DRL, especially under the higher load. This indicates that the VQC-based scheduler learns more effective UE priority scores than the classical MLP actor under the same PMAR budget control, feasibility projection, and safety fallback mechanisms.

\begin{figure}[!t]
\centering
\includegraphics[width=\columnwidth]{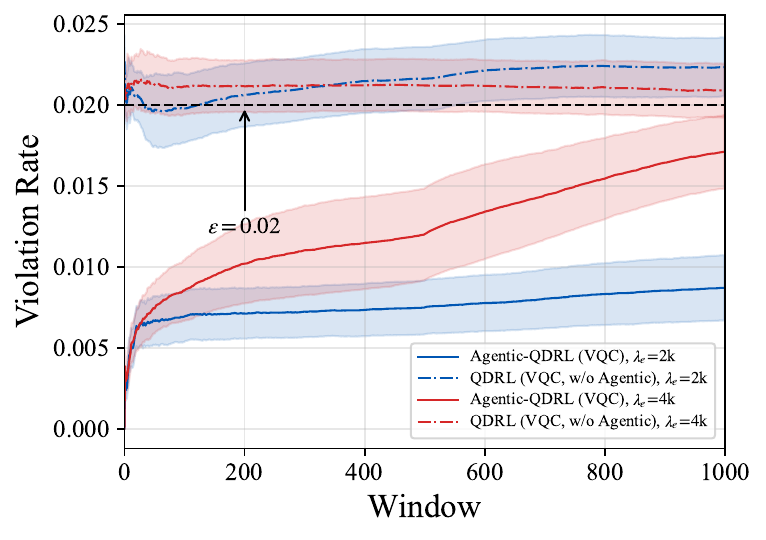}
\caption{URLLC delay-violation rate over slicing windows with and without the agentic PMAR controller under different eMBB loads $\lambda_e$ ($\varepsilon=0.02$).  }
\label{fig:agentic_ablation}
\end{figure}

Fig.~\ref{fig:agentic_ablation} plots the URLLC delay-violation rate over
slicing windows with and without the agentic PMAR controller under two eMBB
traffic loads. We observe that the full Agentic-QDRL keeps the violation rate
below the SLA target for both $\lambda_e=2000$ and $\lambda_e=4000$ bits per
scheduling interval. In contrast, QDRL without the agentic controller stays
close to or above the SLA target because its fixed slice budgets cannot adapt to
the evolving delay-reliability state. The gap between the solid and dashed curves shows the role of slow-time-scale budget adaptation. When the eMBB load increases from $\lambda_e=2000$ to $\lambda_e=4000$, the URLLC delay-violation rate of Agentic-QDRL grows, since heavier eMBB traffic increases contention for the shared PRB pool. However, the PMAR controller allocates sufficient URLLC resources based on the memory-averaged delay-violation rate, keeping the final violation rate below $\varepsilon$. This demonstrates that agentic budget adaptation is essential
for maintaining URLLC delay reliability under changing eMBB load.

\begin{figure}[!t]
\centering
\includegraphics[width=\columnwidth]{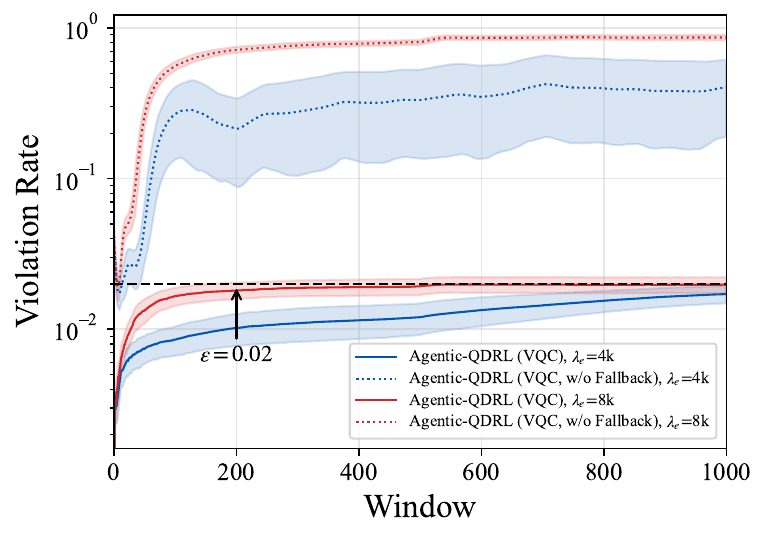}
\caption{URLLC delay-violation rate over slicing windows with and without safety fallback under different eMBB loads $\lambda_e$ ($\varepsilon=0.02$).}
\label{fig:fb_ablation}
\end{figure}

Fig.~\ref{fig:fb_ablation} plots the URLLC delay-violation rate over slicing windows with and without the safety fallback under two heavier eMBB traffic loads, $\lambda_e=4000$ and $\lambda_e=8000$ bits per scheduling interval. We observe that Agentic-QDRL with safety fallback keeps the violation rate below the SLA target for both loads, whereas removing the fallback drives the violation rate far above the target. The no-fallback curves rise to the $10^{-1}$ range, which is more than one order of magnitude above the SLA target under the heavier load. This gap shows the importance of the deterministic EDF/PF safety scheduler. With fallback enabled, unsafe learned allocations are replaced by deadline-aware URLLC scheduling and PF-based eMBB scheduling. Without fallback, the nominal VQC-based allocation is executed directly even when the URLLC operating point becomes unsafe. The degradation is more severe under $\lambda_e=8000$ because heavier eMBB traffic increases PRB contention and reduces service opportunities for urgent URLLC packets. These results show that fast-time-scale safety shielding is essential for maintaining URLLC delay reliability under heavy traffic load.

\begin{figure}[!t]
\centering
\includegraphics[width=\columnwidth]{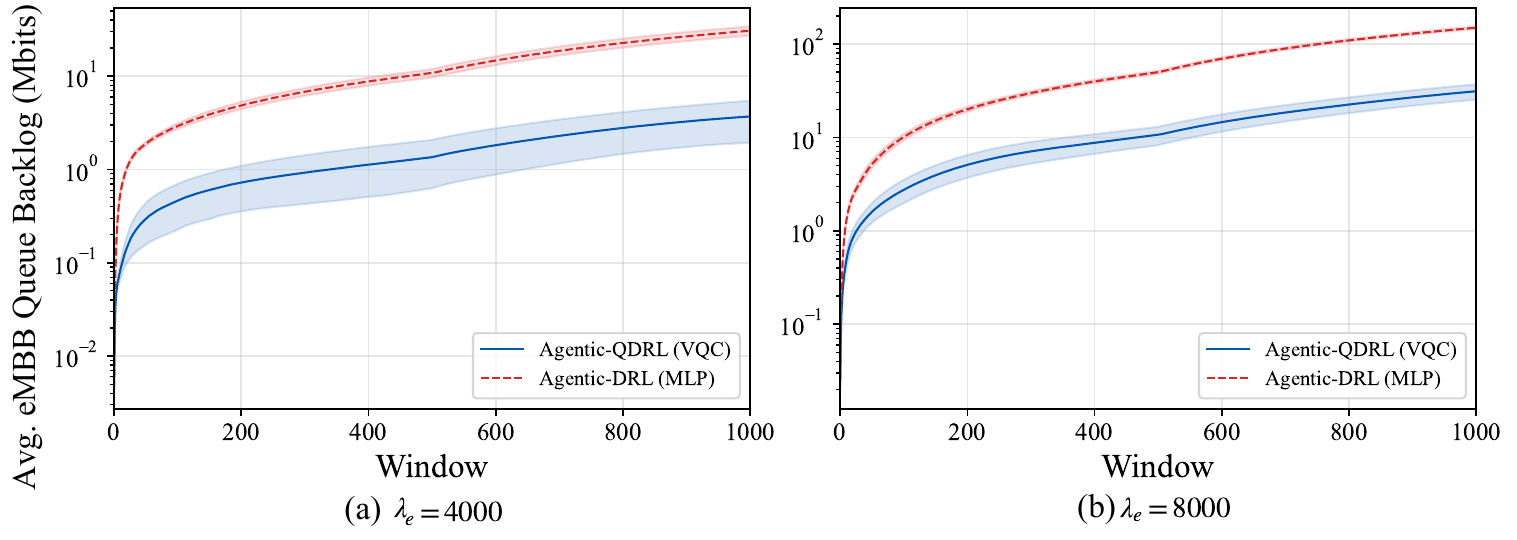}
\caption{Average eMBB buffer backlog over slicing windows for
Agentic-QDRL (VQC) and Agentic-DRL (MLP) under different eMBB traffic loads.}
\label{fig:queue}
\end{figure}

Fig.~\ref{fig:queue} examines eMBB queue accumulation under heavier traffic loads. The average eMBB queue backlog measures how much eMBB traffic remains unserved, and persistent backlog growth indicates increasing congestion. We observe that Agentic-QDRL (VQC) maintains a substantially lower backlog than Agentic-DRL (MLP) under both $\lambda_e=4000$ and $\lambda_e=8000$ bits per scheduling interval. The gap becomes more pronounced under $\lambda_e=8000$, where the Agentic-DRL (MLP) baseline accumulates backlog much faster. This indicates that Agentic-QDRL (VQC) serves eMBB traffic more efficiently and mitigates queue buildup under heavy traffic load.

\begin{figure}[!t]
\centering
\includegraphics[width=\columnwidth]{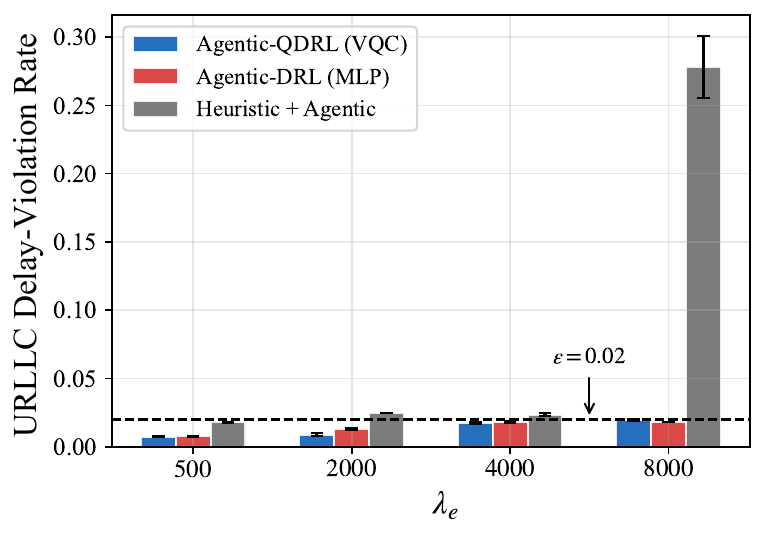}
\caption{URLLC delay-violation rate under various eMBB loads $\lambda_e$ for different algorithms ($\varepsilon=0.02$).}
\label{fig:bar_violation}
\end{figure}

\begin{figure}[!t]
\centering
\includegraphics[width=\columnwidth]{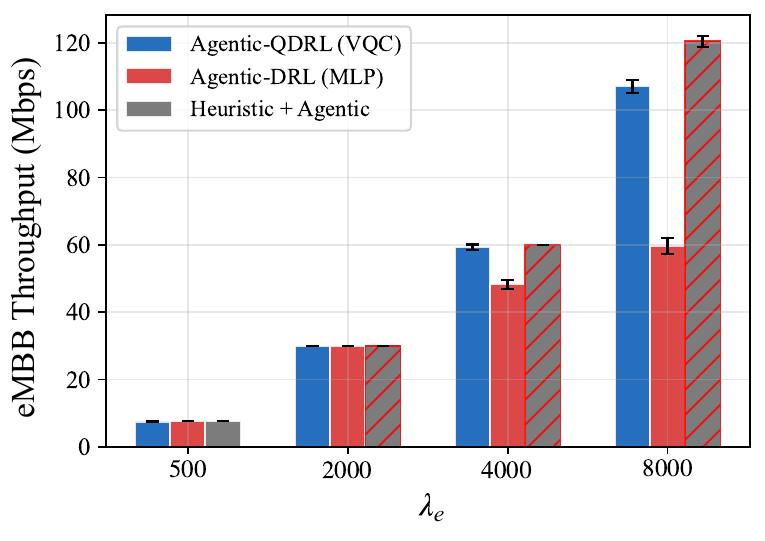}
\caption{eMBB throughput under various eMBB loads $\lambda_e$ for different algorithms. Hatched bars indicate operating points that violate the URLLC SLA target $\varepsilon=0.02$.}
\label{fig:bar_throughput}
\end{figure}

Figs.~\ref{fig:bar_violation} and~\ref{fig:bar_throughput} show the URLLC delay-violation rate and eMBB throughput under different eMBB loads $\lambda_e\in\{500,2000,4000,8000\}$ bits per scheduling interval for different algorithms. At light load, all methods achieve similar throughput and satisfy the SLA target. As the eMBB load increases, the Heuristic+Agentic baseline remains competitive in raw throughput but violates the URLLC SLA target, with its violation rate growing to about $0.28$ at $\lambda_e=8000$, more than one order of magnitude above $\varepsilon$. In contrast, both learned methods keep the violation rate below $\varepsilon$ across all loads, while Agentic-QDRL (VQC) provides a higher feasible throughput than Agentic-DRL (MLP) under heavy load, reaching about $107$~Mbps versus $60$~Mbps at $\lambda_e=8000$. As indicated by the hatched bars in Fig.~\ref{fig:bar_throughput}, the high throughput of the heuristic is obtained at infeasible operating points, whereas Agentic-QDRL (VQC) delivers the best throughput among the SLA-feasible methods. These results show that the proposed quantum-agentic design improves the practical URLLC delay-reliability--eMBB throughput tradeoff.

\section{Conclusion}
In this paper, we considered a URLLC/eMBB downlink OFDMA RAN slicing system, where eMBB throughput maximization needs to be balanced with URLLC delay-reliability requirements. We formulated this tradeoff as a long-term eMBB throughput maximization problem subject to URLLC delay-violation, PRB exclusivity, and slice-budget constraints. To solve this NP-hard problem, we presented Agentic-QDRL, a two-time-scale framework that couples a slow agentic PMAR budget controller with a fast QDRL scheduler, and enforces hard scheduling constraints through deterministic feasibility projection and EDF/PF safety fallback. Simulation results showed that Agentic-QDRL converges to higher eMBB throughput than the classical DRL baseline, maintains a substantially smaller eMBB backlog, and keeps the steady-state URLLC delay-violation rate below the SLA target across all tested loads. Under heavy load, Agentic-QDRL provides the best feasible throughput while the heuristic baseline becomes infeasible. The ablation studies further confirmed that both agentic budget adaptation and safety fallback are essential for maintaining URLLC delay reliability. Future work will extend the framework to multi-cell interference coordination
and finite-shot validation of the VQC scheduler.

\section*{Acknowledgment}
This work was supported by Mitacs, Ericsson Canada, and the NSERC Canada Research Chairs Program.

\bibliographystyle{IEEEtran}
\bibliography{agentic_DRL_slices}

\end{document}